# Unsupervised anomaly detection on cybersecurity data streams: a case with BETH dataset

Evgeniy Eremin

*Abstract* — **In today's world, the importance of cybersecurity for various systems is growing every year. The number of information security events generated by information security tools grows up with the development of the IT infrastructure. At the same time, the cyber threat landscape does not remain constant, and monitoring should take into account both already known attack indicators and those for which there are no signature rules in information security products of various classes yet. Detecting anomalies in large cybersecurity data streams is a complex task that, if properly addressed, can allow for timely response to atypical and previously unknown cyber threats. The possibilities of using of offline algorithms may be limited for a number of reasons related to the time of training and the frequency of retraining. Using stream learning algorithms for solving this task is capable of providing near-real-time data processing. This article examines the results of ten algorithms from three Python stream machine-learning libraries on BETH dataset with cybersecurity events, which contains information about the creation, cloning, and destruction of operating system processes collected using eBPF. The ROC-AUC metric and the total processing time for processing with these algorithms are presented. Several combinations of features and the order of events have been considered. In conclusion, some notes are given about the most promising algorithms, and possible directions for future research are outlined.**

## I. Introduction

Anomaly detection and outlier detection are well-known tasks for which methods of probability and statistics theory, machine and deep learning, and graph theory are used. Ensuring cybersecurity is a modern and complete challenge in a practical field where there are no universal recipes. New classes of products that increase protection from attackers appear periodically. At the same time, attackers are not standing still, and new threats, such as zero-day attacks, make it impossible to fully rely on signature detection tools to identify malicious behavior.

As society becomes digital, the number of systems that need to be secured at an acceptable level increases. Data streams from information protection systems can reach hundreds of thousands and millions of events per second (EPS). In such conditions, it becomes impossible not to use automatic intelligent analysis of incoming data.

The feature of systems that analyze security events is the need to minimize the time of detection of illegitimate events. From the point of view of information security monitoring, one of the key metrics of performance is the mean time to detect (MTTD). In fact, to minimize this metric, event analysis system must operate in near real-time mode.

The most impressive results in detecting anomalies are achieved by methods [1], which can be challenging to implement in practice for processing large data streams in near real-time.

This work considers the processing of the cybersecurity event stream of the Linux operating system kernel deployed in a cloud infrastructure. The analysis of such data streams makes it possible to solve in practice the task of monitoring the security of container orchestration systems in the cloud. Such events can be collected and analyzed by various types of systems, including Endpoint Detection and Response (EDR), Security Information and Event Management (SIEM), and User and Entity Behavioral Analytics (UEBA).

The rest of this paper is organized as follows. Section II provides a brief review of relevant literature related to this work. Used dataset and algorithms presented in Section III. Methodology of evaluation experiments provided in Section IV. In Section V the results of experiments are listed and some explanations provided. Conclusion of article and possible future work presented in Section VI.

## II. Related Work

When considering the task of unsupervised anomaly detection on cybersecurity data streams, it is possible to identify several relevant areas of work.

First, there is work on the application of algorithms for detecting anomalies in data streams. An overview of these algorithms is provided in [2]. In [3], a scalable real-time system for streaming cybersecurity logs based on the use of the Spark framework and the MLlib machine-learning library is considered. In [4], a comparison of various methods for extracting features from a stream of data collected from network traffic analysis devices was conducted. In [5], the use of deep neural networks for insider threat detection in streaming data is discussed.

Secondly, it is the works on the use of multidimensional data clustering algorithms in the context of the cybersecurity domain. A detailed analysis and review of the use of clustering algorithms in UEBA systems can be found in [6].

The third area of work can be identified as addressing the specific issue of anomaly detection within system logs. The work [7] presents the use of fine-tuned language models for solving the task of anomaly detection in log data. The article

E.O. Eremin is with RTU MIREA, Russia, Moscow
(e-mail: e.o.eremin@gmail.com).

[8] discusses the use of a fuzzy CNN autoencoder for this task, using the datasets HDFS1, BGL, and Villani.

Finally, the detection of abnormal processes based on events generated by the extended Berkeley Packet Filter (eBPF) is discussed in the paper [9]. The BETH dataset is used in a number of papers, including [10], [11], [12] and [13]. In the original study that introduced the dataset [10], several non-streaming methods were chosen as baselines, including Isolation Forest, Robust Covariance, One-Class Support Vector Machine (SVM), and Variational Autoencoder with State Density Estimation. The best ROC-AUC score for the test dataset was achieved by Isolation Forest, with a score of 0.850. In the work [11], the usage of graph neural networks with embeddings from transformers is proposed. The best results were achieved by the GraphSAGE-128 + IForest model trained on the T5-VAE embedding with ROC-AUC of 0.932 for SUS labels and 0.951 for EVIL labels.

Unlike the previously mentioned works, this study focuses on the use of streaming algorithms for detecting anomalies in the BETH dataset, which contains information about system processes. The main idea of the work is to test the hypothesis about the possibility of using such algorithms and to compare the results with those obtained using offline methods for anomaly detection.

## III. Background

### A. eBPF

Extended Berkley Packet Filter is a technology for launching applications in the Linux kernel space. It is more secure than using the Linux kernel module mechanism. In recent years, the use of this technology has been growing due to its capabilities in terms of security, observability and tracing [14]. In particular, many companies use eBPF to monitor the security of Kubernetes containers.

### B. Unsupervised stream learning

In the context of the possibility of detecting anomalies in real time on a sufficiently large stream of events, the number of methods suitable for this task is limited. Supervised methods require the preliminary marking of data, which is very difficult to obtain in real time. At the same time, unsupervised methods can be divided into offline, semi-online and online [2]. Offline methods are able to extract more information from the data, but are less resistant to changes in newly incoming data, such as shift, change in distribution, etc. The use of offline learning methods requires periodic full retraining of models, which can be a difficult engineering task in the case of large volumes of processed data streams. Semi-online and online methods allow for data changes to be taken into account and reduce the time spent on training, but they may be less effective in terms of accuracy when detecting anomalies.

In this work, ten algorithms of semi-online and online learning are used, the implementations of which are available in modern libraries for stream machine learning River [15], PySAD [16] and StreamAD [17]:

1) Half-Space Tree – online version of isolated trees. In this work used implementation from River library;
2) IForestASD. The idea of this algorithm is to use sliding windows of a predetermined size, within which the original isolation forest is applied and the anomaly score [18] is calculated for events within the window. In this work, the implementation from the PySAD library was used, which has no concept drift detection. The implementation is based on pod.models.forest from the PyOD library [19];
3) Incremental Local Outlier Factor. This is online version of the Local Outlier Factor (LOF) [20]. Main idea of this method is to identify outliers based on density of local neighbors. Implementation from River library was used;
4) KitNet. This algorithm is based on the idea of using small autoencoders trained to imitate (reconstruct) patterns in incoming data, whose performance improves during operation. The main limitation of this method is that it needs to be trained on normal data [21];
5) LODA. Ensemble of weak anomaly detectors – one-dimensional histograms for approximation the probability density of input data [22]. In this work used implementation from StreamAD library;
6) Stochastic implementation of One-Class Support Vector Machine from River library, not exactly matched with its batch formulation;
7) Robust Random Cut Forest – dynamically maintained Robust Random Cut Trees. It differs from an isolated forest in that the dimension to cut has chosen uniformly at random [23]. In this work used implementation from StreamAD library;
8) RS-Hash – Lightweight subspace outlier algorithm based on randomized hashing to score data points and has subspace interpretation [24]. In this work used implementation from StreamAD library;
9) Storm (Exact-STORM) – method with sliding windows and distance-based anomaly scores [25]. In this work used implementation from PySAD library;
10) xStream – Density-based ensemble method that can work on row streams and feature-evolving streams [26]. In this work used implementation from StreamAD library.

### C. BETH Cybersecurity Dataset

This dataset was made publicly available in 2021. It contains events from eBPF-based sensors that log the creation, cloning, and termination of processes, as well as network traffic events, mainly DNS requests. In this work, as in several others that use this dataset, only process monitoring events are used [11] – [13]. Each event contains 14 raw features, 9 of which are numeric. Not all raw features are used in this work; in particular, arguments are not used due to the structure, which is difficult to process in a stream.

Train sample contains events from 8 hosts, validation sample contains events from 4 hosts and testing sample contains events from one host. Because of this we consider one host-independent model.

The training and validation datasets do not include events that occurred during an attack. Events in the dataset are marked with two tags – EVIL and SUS. EVIL – events of processes that are clearly related to an attack. Such events are only in the test sample. SUS – non-typical events for processes that are not unequivocally malicious, but still quire the attention from a security analyst. In this work, tags are only used to calculate the final metrics.

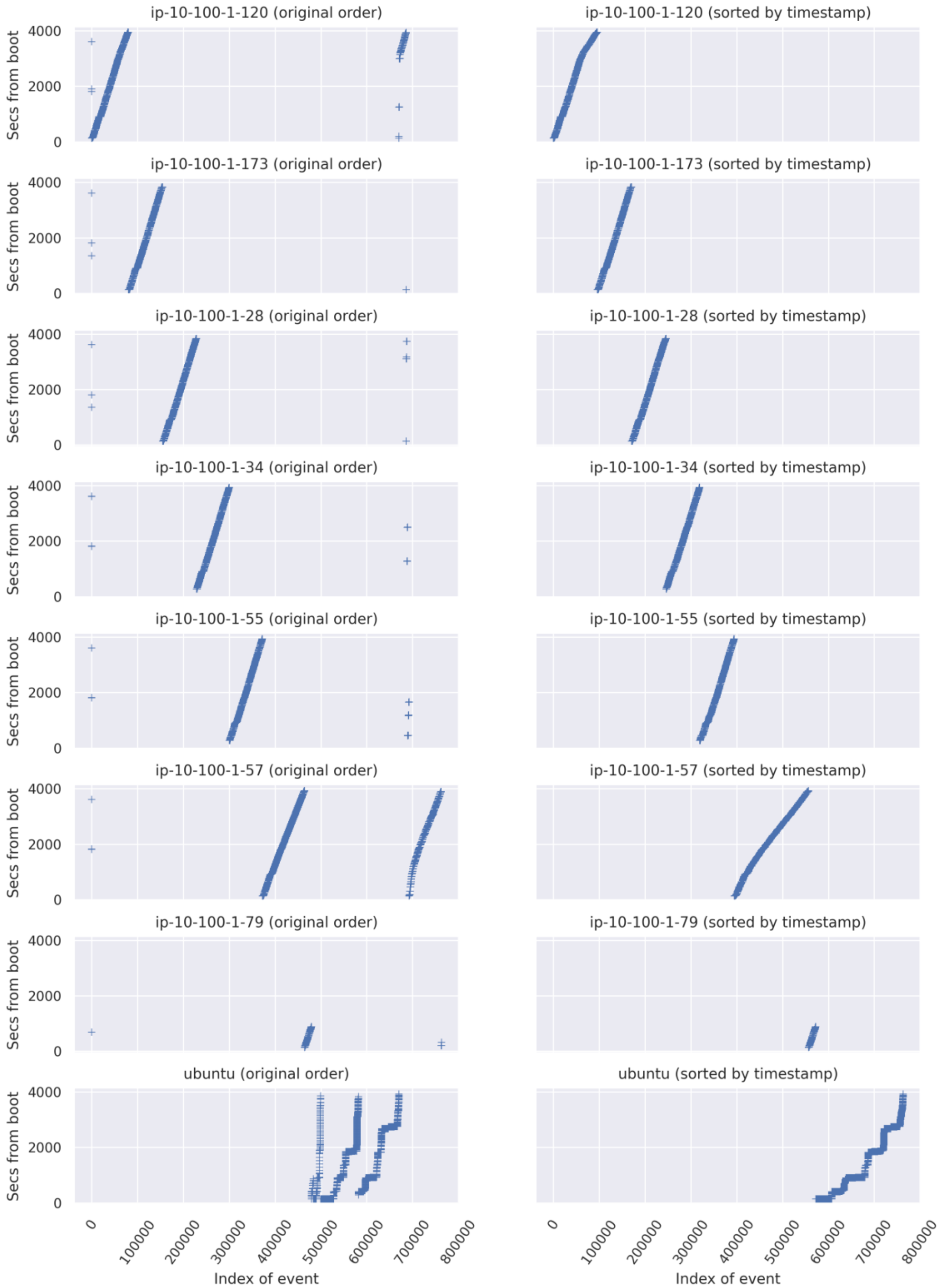

Fig. 1. Events order in train split of BETH dataset. Left column – original order. Right column – after sorting by timestamp and host

## IV. Methodology

### A. Data Preprocessing

As in the original article presenting the BETH dataset, additional features are used:

- processId_nonOS — binary flag that determines whether the process is spawned by the user or the OS;
- parentProcessId_nonOS — the same as previous flag, but for parent process;
- userId_nonOS — binary flag that determines whether the account is a user account or a system account;
- returnValue_error — process finished with or without error.

Unlike the code that comes with the original article, attributes are added to existing event fields. The event names of the parent process are enriched where possible. Process arguments are not used. The string values are encoded in a stream mode using the Ordinal Encoder from RiverML library.

Network logs (DNS) are not utilized in this work.

For the OCSVM model, StandardScaler is used according to the recommendations in the RiverML library documentation, and MinMaxScaler is used for HSTree.

Since the timestamp field contains the number of seconds since the OS was loaded, it is not possible to reliably determine the chronological order of events from different OS sessions. However, in the case of streaming training, the order of events at the input of the model can affect its predictions. Therefore, in the train data, the events were also sorted by timestamp and hostname. The charts of the timestamp dependencies on the event number in the sample are shown in Figure 1. The charts of the events in their original order are located on the left, and the sorted charts are on the right. All of the models considered process the data in both orders for subsequent comparison.

### *B. Evaluation and metrics*

All experiments were conducted on Ubuntu 22.04 LTS equipped with Intel® Core™ i7-7700 at 3.6 GHz and 48 GB DDR4 memory. For programming, Python 3.11 was used. The training and testing samples were used for training and calculating anomaly scores. At the same time, the evaluation metrics were calculated only for the testing sample, separately for the EVIL and SUS labels. The validation sample was not used in the experiments. All models were initialized using the default hyperparameters provided by their implementation in the respective libraries. The experiments were conducted for various combinations of features and data order:

- original or sorted order by hostname and timestamp;
- enriched event by parent process name or not.

Despite the fact that not all the models discussed are stochastic, each model is evaluated using five different random seeds and the results are averaged.

The metrics chosen are ROC-AUC, as in the original work, and the total processing time of the entire dataset by the model.

Although there are both opinions for using ROC-AUC on unbalanced datasets (as in the case of BETH) and against, other binary classification metrics such as Accuracy, Precision, Recall and F1 are not used in this work. This is because the ROC curve allows you to choose a threshold that meets a specific balance between True Positive Rate (TPR) and False Positive Rate (FPR). The optimal TPR and FPR ratio can be chosen from the ROC curve; however, in the context of analyzing suspicions of information security incidents with a specific intelligent tool, minimizing false positives or maximizing detected anomalies may be of greater importance. Accordingly, depending on the selected balance of TPR and FPR, binary classification metrics will take different values, which is why they are not presented.

The ROC-AUC calculation is performed after processing the entire test sample using roc_auc_score from the scikit-learn library [27]. The River library has the ability to calculate metrics on the fly, but this feature is not used because it is more error-prone than calculating metrics after the fact.

The total processing time is calculated by adding up the processing times for each event in the training and testing splits. Fit and score are then performed on each event. Preliminary processing of the additional feature and enrichment of the parent process is performed separately and not included in the measurements. This is related to the desire to provide convenience for repeated runs, so that we can use the feature sets that have already been prepared for the training and testing samples.

## V. RESULTS AND DISCUSSION

The numerical results of the experiments are presented in Table 1 for the anomalies marked as EVIL and in Table 2 for the anomalies marked as SUS.

For HSTree, LODA, and RRCF models, ROC-AUC values were less than 0.5. Inverting anomaly scores from these methods gives ROC-AUC values between approximately 0.8 and 1.0. This means that these models confuse anomalies and typical events vice versa. HSTree confuses it across different seeds, which indicates the poor capabilities of this model in this particular case. The best results for ROC-AUC belong to KitNet, ILOF, and RS-Hash.

The fastest processing speed for the entire dataset was achieved by OCSVM, KitNet, HSTree, and ILOF, with processing times of more than 2,500 EPS.

It can be seen that, on the used dataset, some anomaly detection algorithms outperform the results presented in the ROC-AUC metric in the works of [10] - [12]. In this case, only basic feature engineering and default hyperparameter values of all discussed models are used. In the paper [13], supervised methods were used and the results on ROC-AUC were not presented, so a comparison could not be made.

All algorithms, except Storm, were sensitive to the order of events in which incremental training takes place. Unfortunately, there is not a clear connection between the quality of the detection and the results. Therefore, for the KitNet algorithm, sorting the training sample by timestamp improves the result when detecting anomalies labeled as EVIL, but it leads to a decrease in the performance metric for events labeled as SUS.

Enrichment of the parent process name with subsequent processing by the Ordinal Encoder stream for some algorithms worsens the results, and for some it improves, and for different labels (EVIL and SUS) the situation with the same algorithm can be the opposite (for example, the same KitNet).

The best ROC-AUC metric in most cases was the KitNet algorithm, which is due to:

1) Absence of events with the EVIL label in the training sample;
2) Small number of events with the SUS label in the training sample (0.02% of the total number).

«Contamination» of the training sample with SUS and EVIL events (which is what we can observe in real practice of monitoring) would lead to a decrease in the result shown.

All of the above observations allow us to conclude that even the most promising of the algorithms reviewed, despite their high discriminatory ability, are strongly dependent on the processed data and their results should be evaluated by a security analyst.

TABLE I
MODELS PERFORMANCE ON EVIL LABELS

| Model | Sorted | Enriched | Time, seconds | ROC-AUC |
|---|---|---|---|---|
| HSTree | no | no | 215.254±9.094 | 0.216±0.412 |
| | | yes | 212.872±15.805 | 0.052±0.007 |
| | yes | no | 209.177±7.672 | 0.215±0.422 |
| | | yes | 216.206±6.950 | 0.042±0.012 |
| IForestASD | no | no | 1865.587±11.517 | 0.710±0.013 |
| | | yes | 1872.442±12.069 | 0.676±0.009 |
| | yes | no | 1860.352±10.495 | 0.710±0.014 |
| | | yes | 1870.231±12.328 | 0.719±0.015 |
| ILOF | no | no | 374.580±2.544 | 0.880±0.000 |
| | | yes | 384.471±3.058 | **0.978**±0.000 |
| | yes | no | 242.972±1.499 | 0.712±0.000 |
| | | yes | 384.452±3.849 | 0.883±0.000 |
| KitNet | no | no | 176.138±0.503 | 0.933±0.000 |
| | | yes | 274.922±2.356 | 0.933±0.000 |
| | yes | no | 172.029±0.928 | **0.994**±0.000 |
| | | yes | 224.652±1.549 | **0.982**±0.000 |
| LODA | no | no | 816.395±5.402 | 0.097±0.000 |
| | | yes | 821.759±5.795 | 0.149±0.001 |
| | yes | no | 817.694±5.824 | 0.097±0.000 |
| | | yes | 818.895±5.378 | 0.140±0.001 |
| OCSVM | no | no | **53.823**±0.665 | 0.659±0.000 |
| | | yes | **52.928**±0.695 | 0.676±0.000 |
| | yes | no | **55.409**±0.486 | 0.711±0.000 |
| | | yes | **50.643**±0.377 | 0.572±0.000 |
| RRCF | no | no | 1742.404±75.728 | 0.047±0.001 |
| | | yes | 1573.353±11.010 | 0.044±0.000 |
| | yes | no | 1452.139±51.779 | 0.047±0.002 |
| | | yes | 1408.911±9.375 | 0.045±0.001 |
| RS-Hash | no | no | 415.451±4.079 | **0.967**±0.001 |
| | | yes | 437.649±0.792 | 0.955±0.000 |
| | yes | no | 417.394±3.099 | 0.966±0.001 |
| | | yes | 438.476±1.142 | 0.955±0.001 |
| Storm | no | no | 3349.882±102.540 | 0.932±0.000 |
| | | yes | 3313.516±83.078 | 0.931±0.000 |
| | yes | no | 3362.405±184.883 | 0.932±0.000 |
| | | yes | 3297.809±78.910 | 0.931±0.000 |
| xStream | no | no | 680.401±3.902 | 0.917±0.009 |
| | | yes | 701.267±2.618 | 0.918±0.003 |
| | yes | no | 681.296±3.363 | 0.916±0.010 |
| | | yes | 702.592±2.585 | 0.917±0.003 |

ROC-AUC less than 0.5 means that model confuses anomalous and benign events. Best values highlighted by bold font.

## VI. CONCLUSION AND FUTURE WORK

Some of discussed incremental models (like KitNet, ILOF, RS-Hash, xStream) can be used for anomaly detection in cybersecurity data streams “as-is”, i.e. with minimal feature engineering, default hyper-parameters and high EPS processing rate. Detections from these models can be used to assist security analysts in their operational activities. Advanced feature engineering and tuning hyper-parameters may give better results.

Future directions of work:

1) Reviewing performance of batch-based learning models for this task. Batch learning can be more difficult in engineering meaning while using in real product environment, but can find more complex dependencies in data;
2) Advanced feature engineering, such as for process arguments, can be done using arrays. Strings and lists of strings can be represented as embedding vectors derived from a domain-specific language model. Methods of unsupervised feature selection on streaming data also could be used;
3) Entity-based models (for certain host, certain process name, etc.) can be interesting for using in particular infrastructure;
4) Evaluating various models on other cybersecurity streaming data, e.g. Sysmon events, etc.;
5) Evaluating more deep learning models, including graph neural networks.

TABLE II
MODELS PERFORMANCE ON SUS LABELS

| Model | Sorted | Enriched | Time, seconds | ROC-AUC |
|---|---|---|---|---|
| HSTree | no | no | 215.254±9.094 | 0.235±0.390 |
| | | yes | 212.872±15.805 | 0.081±0.005 |
| | yes | no | 209.177±7.672 | 0.222±0.391 |
| | | yes | 216.206±6.950 | 0.083±0.014 |
| IForestASD | no | no | 1865.587±11.517 | 0.671±0.016 |
| | | yes | 1872.442±12.069 | 0.678±0.006 |
| | yes | no | 1860.352±10.495 | 0.679±0.016 |
| | | yes | 1870.231±12.328 | 0.706±0.019 |
| ILOF | no | no | 374.580±2.544 | 0.832±0.000 |
| | | yes | 384.471±3.058 | 0.950±0.000 |
| | yes | no | 242.972±1.499 | 0.662±0.000 |
| | | yes | 384.452±3.849 | 0.771±0.000 |
| KitNet | no | no | 176.138±0.503 | **0.990**±0.000 |
| | | yes | 274.922±2.356 | **0.983**±0.000 |
| | yes | no | 172.029±0.928 | **0.976**±0.000 |
| | | yes | 224.652±1.549 | **0.953**±0.000 |
| LODA | no | no | 816.395±5.402 | 0.138±0.001 |
| | | yes | 821.759±5.795 | 0.188±0.001 |
| | yes | no | 817.694±5.824 | 0.138±0.001 |
| | | yes | 818.895±5.378 | 0.170±0.001 |
| OCSVM | no | no | **53.823**±0.665 | 0.743±0.000 |
| | | yes | **52.928**±0.695 | 0.766±0.000 |
| | yes | no | **55.409**±0.486 | 0.765±0.000 |
| | | yes | **50.643**±0.377 | 0.731±0.000 |
| RRCF | no | no | 1742.404±75.728 | 0.093±0.001 |
| | | yes | 1573.353±11.010 | 0.092±0.001 |
| | yes | no | 1452.139±51.779 | 0.094±0.002 |
| | | yes | 1408.911±9.375 | 0.091±0.001 |
| RS-Hash | no | no | 415.451±4.079 | 0.938±0.002 |
| | | yes | 437.649±0.792 | 0.927±0.001 |
| | yes | no | 417.394±3.099 | 0.937±0.003 |
| | | yes | 438.476±1.142 | 0.927±0.002 |
| Storm | no | no | 3349.882±102.540 | 0.914±0.000 |
| | | yes | 3313.516±83.078 | 0.913±0.000 |
| | yes | no | 3362.405±184.883 | 0.914±0.000 |
| | | yes | 3297.809±78.910 | 0.913±0.000 |
| xStream | no | no | 680.401±3.902 | 0.882±0.011 |
| | | yes | 701.267±2.618 | 0.881±0.007 |
| | yes | no | 681.296±3.363 | 0.880±0.012 |
| | | yes | 702.592±2.585 | 0.881±0.007 |

ROC-AUC less than 0.5 means that model confuses anomalous and benign events. Best values highlighted by bold font.

## DATA AVAILABILITY

Source code for reproducing experiments is available at https://github.com/ev-er/unsupervised_stream_ad_beth

BETH dataset is available at https://www.kaggle.com/datasets/katehighnam/beth-dataset

## REFERENCES

[1] Bouman, R., Bukhsh, Z., & Heskes, T. (2024). Unsupervised anomaly detection algorithms on real-

world data: how many do we need?. Journal of Machine Learning Research, 25(105), 1-34.

[2] Lu, T., Wang, L., & Zhao, X. (2023). Review of Anomaly Detection Algorithms for Data Streams. Applied Sciences, 13(10), 6353.

[3] Sánchez-Zas, C., Larriva-Novo, X., Villagrá, V. A., Rodrigo, M. S., & Moreno, J. I. (2022). Design and Evaluation of Unsupervised Machine Learning Models for Anomaly Detection in Streaming Cybersecurity Logs. Mathematics, 10(21), 4043.

[4] Heigl, M., Weigelt, E., Fiala, D., & Schramm, M. (2021). Unsupervised Feature Selection for Outlier Detection on Streaming Data to Enhance Network Security. Applied Sciences, 11(24), 12073.

[5] Tuor, A., Kaplan, S., Hutchinson, B., Nichols, N., & Robinson, S. (2017, February). Deep Learning for Unsupervised Insider Threat Detection in Structured Cybersecurity Data Streams. In AAAI Workshops (pp. 224-231).

[6] Artioli, P., Maci, A., & Magrì, A. (2024). A comprehensive investigation of clustering algorithms for User and Entity Behavior Analytics. Frontiers in big Data, 7, 1375818.

[7] Almodovar, C., Sabrina, F., Karimi, S., & Azad, S. (2024). LogFiT: Log anomaly detection using fine-tuned language models. IEEE Transactions on Network and Service Management, 21(2), 1715-1723.

[8] Gorokhov, O., Petrovskiy, M., Mashechkin, I., & Kazachuk, M. (2023). Fuzzy CNN Autoencoder for Unsupervised Anomaly Detection in Log Data. Mathematics, 11(18), 3995.

[9] Kotenko, I. V., Melnik, M. V., & Abramenko, G. T. (2024, June). Anomaly Detection in Container Systems: Using Histograms of Normal Processes and an Autoencoder. In 2024 IEEE 25th International Conference of Young Professionals in Electron Devices and Materials (EDM) (pp. 1930-1934). IEEE.

[10] Highnam, K., Arulkumaran, K., Hanif, Z., & Jennings, N. R. (2021). Beth dataset: Real cybersecurity data for unsupervised anomaly detection research. In CEUR Workshop Proc (Vol. 3095, pp. 1-12).

[11] Lakha, B., Mount, S. L., Serra, E., & Cuzzocrea, A. (2022, December). Anomaly detection in cybersecurity events through graph neural network and transformer based model: A case study with beth dataset. In 2022 IEEE International Conference on Big Data (Big Data) (pp. 5756-5764). IEEE.

[12] Sushmakar, N., Oberoi, N., Gupta, S., & Arora, A. (2022, June). An unsupervised based enhanced anomaly detection model using features importance. In 2022 2nd International Conference on Intelligent Technologies (CONIT) (pp. 1-7). IEEE.

[13] Khan, L. P., Hossain, A., & Dey, S. (2023, February). Anomaly Detection for Beth Dataset Using Machine Learning Approaches. In 2023 Fifth International Conference on Electrical, Computer and Communication Technologies (ICECCT) (pp. 1-6). IEEE.

[14] Security Observability with eBPF, Natália Réka Ivánkó and Jed Salazar, O'Reilly, 2022

[15] Montiel, J., Halford, M., Mastelini, S. M., Bolmier, G., Sourty, R., Vaysse, R., Bifet, A. (2021). River: machine learning for streaming data in python. Journal of Machine Learning Research, 22(110), 1-8.

[16] Yilmaz, S. F., & Kozat, S. S. (2020). PySAD: A streaming anomaly detection framework in python. arXiv preprint arXiv:2009.02572.

[17] Xu, J., Lin, C., Liu, F., Wang, Y., Xiong, W., Li, Z., ... & Xie, G. (2023). StreamAD: A cloud platform metrics-oriented benchmark for unsupervised online anomaly detection. BenchCouncil Transactions on Benchmarks, Standards and Evaluations, 3(2), 100121.

[18] Ding, Z., & Fei, M. (2013). An anomaly detection approach based on isolation forest algorithm for streaming data using sliding window. IFAC Proceedings Volumes, 46(20), 12-17.

[19] Zhao, Y., Nasrullah, Z., & Li, Z. (2019). Pyod: A python toolbox for scalable outlier detection. Journal of machine learning research, 20(96), 1-7.

[20] Pokrajac, D., Lazarevic, A., & Latecki, L. J. (2007, March). Incremental local outlier detection for data streams. In 2007 IEEE symposium on computational intelligence and data mining (pp. 504-515). IEEE.

[21] Mirsky, Y., Doitshman, T., Elovici, Y., & Shabtai, A. (2018). Kitsune: an ensemble of autoencoders for online network intrusion detection. arXiv preprint arXiv:1802.09089.

[22] Pevný, T. (2016). Loda: Lightweight on-line detector of anomalies. Machine Learning, 102, 275-304.

[23] Guha, S., Mishra, N., Roy, G., & Schrijvers, O. (2016, June). Robust random cut forest based anomaly detection on streams. In International conference on machine learning (pp. 2712-2721). PMLR.

[24] Sathe, S., & Aggarwal, C. C. (2016, December). Subspace outlier detection in linear time with randomized hashing. In 2016 IEEE 16th International Conference on Data Mining (ICDM) (pp. 459-468). IEEE.

[25] Angiulli, F., & Fassetti, F. (2007, November). Detecting distance-based outliers in streams of data. In Proceedings of the sixteenth ACM conference on Conference on information and knowledge management (pp. 811-820).

[26] Manzoor, E., Lamba, H., & Akoglu, L. (2018, July). XStream: Outlier detection in feature-evolving data streams. In Proceedings of the 24th ACM SIGKDD International Conference on Knowledge Discovery & Data Mining (pp. 1963-1972).

[27] Pedregosa, F., Varoquaux, G., Gramfort, A., Michel, V., Thirion, B., Grisel, O., ... & Duchesnay, É. (2011). Scikit-learn: Machine learning in Python. The Journal of machine Learning research, 12, 2825-2830.

**Evgeniy Olegovich Eremin**, Cand.Sc. (Engineering)
ORCID: 0009-0001-4330-2741
Associate professor at the Department of Radioelectronic Systems and Complexes, Institute of Radioelectronics and Informatics, RTU MIREA.
Senior analyst of security event collection and analysis systems, BI.ZONE.